\documentclass{iaus}
\usepackage{graphicx}

\title[Masers and star formation]{Masers and star formation}

\author[Fish]{Vincent L.~Fish}

\affiliation{Jansky Fellow, National Radio Astronomy Observatory, 1003
  Lopezville Rd., Socorro, NM 87801, USA \break email:
  \texttt{vfish@nrao.edu}}

\pubyear{2007}
\volume{242}
\date{\today}
\jname{Astrophysical masers and their environments}
\editors{J.~Chapman \& W.~Baan, eds.}
\begin{document}

\maketitle

\begin{abstract}
Recent observational and theoretical advances concerning astronomical
masers in star forming regions are reviewed.  Major masing species are
considered individually and in combination.  Key results are
summarized with emphasis on present science and future prospects.
\end{abstract}

\firstsection
\section{Introduction} \label{intro}

This review summarizes maser results pertinent to star formation
appearing in the literature since the last maser meeting (IAU
Symposium 206).  References are drawn from recent literature when
possible.

\section{Masing species} \label{masers}

\subsection{Water (H$_2$O)} \label{h2o}

The 22.235 GHz water line is the predominant water maser line.  Masers
in this transition are very bright, easily observable, and inverted
under a wide range of conditions (e.g., \cite{Babkovskaia04}).
Several millimeter and submillimeter transitions of water are also
seen as masers.  Discussion of these transitions can be found in the
section of these proceedings devoted to millimeter and submillimeter
masers.

Water masers are frequently seen in outflows from both high-mass and
low-mass YSOs (\cite{Honma05}; \cite{Moscadelli05}; \cite{Goddi06a};
\cite{Moscadelli06}).  These jets are seen in deceleration
(\cite{Imai02}) and often have substructure on AU scales
(\cite{Torrelles03}; \cite{Furuya05}; \cite{Uscanga05}).

Water masers are sometimes believed to trace disks as well as outflows
(\cite{Seth02}; \cite{Gallimore03}), possibly excited by an expanding
shock wave.  Indeed, shocks likely are responsible for arc-like maser
distributions (\cite{Honma04}) and may excite masers in accreting
material as well (\cite{Menten04}).  Water masers appear in Bok
globules (\cite{Gomez06}), likely tracing bipolar molecular outflows
(\cite{dGm06}), as well as bright rimmed clouds (\cite{Valdettaro05}),
again likely associated with outflows (\cite{Urquhart06}).  The
location of water masers near the ionization front of large
H~\textsc{ii} regions provides evidence in support of triggered star
formation (\cite{Healy04}).  The common thread of all these
environments is the existence of energetic shocks, which fits with
conventional wisdom regarding water maser pumping.

Despite the small Zeeman splitting coefficient of water, line-of-sight
magnetic fields of tens to hundreds of milligauss have been measured
in star forming regions via water maser Zeeman splitting
(\cite{Sarma02}; \cite{Vlemmings06b}), although direct interpretation
of the Stokes V profile as a magnetic field strength may be in error
by up to a factor of two depending on local velocity and magnetic
field gradients (\cite{Vlemmings06a}).  Linear polarization
observations of water masers can provide information on the
orientation of the magnetic field in the plane of the sky.  The
hourglass morphology of the magnetic field in W3 IRS 5 appears to be
due to the processes of collapse rather than a result of the outflow
traced by water masers (\cite{Imai03}).  Further interpretation of
water maser polarization can be found in the review by Wouter
Vlemmings.

Water masers have a fractal distribution over 4 orders of magnitude in
spatial scale, possibly indicating that they appear at the turbulent
dissipation scale (\cite{Strelnitski02}; \cite{Ripman06}; see also
Vladimir Strelnitski's contribution to these proceedings).  Turbulence
may also be responsible for variability, including changes in the
line-of-sight velocity of individual components, of the water masers
in some sources (\cite{Lekht06a}).  Larger-scale variations may
contribute as well, such as changes in outflow parameters or cyclic
variability of the central star (\cite{Pashchenko05};
\cite{Lekht06b}).  An ordered structure is inferred in W31(2) from
successive flaring of features at different velocities
(\cite{Lekht05}).

\subsection{Methanol (CH$_3$OH)} \label{methanol}

Methanol masers divide into two categories, known as class I and class
II, based on their propensity for certain transitions to produce
masers while others are seen in absorption.  Traditionally, class I
and class II masers do not mix (e.g., \cite{Ellingsen05}); however,
fine-tuned conditions may rarely excite lines from both classes
simultaneously, as appears to be the case in OMC-1
(\cite{Voronkov05}).  For purposes of this review, class I and class
II methanol masers will be treated separately.  Improved laboratory
data for rest frequencies have been obtained for many methanol
transitions in both classes (\cite{Muller04}).

\subsubsection{Class I}

Class I masers are primarily collisionally pumped.  They are typically
found in younger sources that are Class II masers and may trace
distant parts of outflows interacting with dense molecular gas
(\cite{Beuther05}; \cite{Ellingsen06}).  Comparison of interferometric
maps of the 9.9 and 104.3~GHz transitions with H$_2$ data confirms the
outflow association in IRAS 16547$-$4247 (\cite{Voronkov06}).  Linear
polarization suggests that Class I masers may appear in oblique shocks
parallel to the outflow axis and perpendicular to the magnetic field
in OMC-2 (\cite{Wiesemeyer04}), although interpretation of methanol
polarization may be complicated (e.g., \cite{Elitzur02}).

Many different Class I transitions have been observed.  Weak masers at
84.5 and 95.2~GHz to the southwest of W3(OH) provide strong evidence
in support of collisional pumping and allow for physical conditions to
be inferred (\cite{Sutton04}).  The latter transition is commonly seen
as a maser in both Class I and Class II sources (\cite{Minier02}).
Strong 36~GHz maser emission is believed to be an indicator of an
early evolutionary stage, as may line ratios of other transitions
(\cite{Gillis05}; \cite{Hoffmann06}).  The intensity ratio of
highly-excited 146.6 and 156.8~GHz methanol masers may be a sensitive
probe of density and temperature in masing regions
(\cite{Lemonias06}).  Short-timescale variability is seen in the 44
and 146.6~GHz transitions (\cite{Pratap06}).

\subsubsection{Class II}

The key Class II maser transitions are at 6668 and 12178~MHz.  Class
II masers are often found in an earlier evolutionary stage than
ultracompact (UC) H~\textsc{ii} regions (e.g., \cite{Minier05}), but
observations of methanol masers cospatial with both millimeter and
centimeter continuum emission indicate that Class II masers appear
over a wide range of evolutionary stages (\cite{Pestalozzi06}).  While
the lower stellar mass limit for methanol masers is still a subject of
research, 6.7~GHz masers do not appear below approximately 3 solar
masses (\cite{Minier03}).

Linear structures of masers with organized velocity structures have
led some to conclude that methanol masers often trace edge-on disks
(\cite[Norris \etal\ 1993, 1998]{Norris93,Norris98}).  Observations of
maser proper motions (\cite{Minier00}), shocked H$_2$
(\cite{deBuizer03}), and SiO (\cite{deBuizer06b}) indicate that, in
the majority of cases at least, methanol masers are aligned with an
outflow, not a disk.  These structures may be explained by propagation
of a shock front into a region with large-scale velocity structure,
such as rotation (\cite{Dodson04}).  Disk candidate sources remain
(\cite{Slysh02b}; \cite{Pestalozzi04}; \cite{Pillai06}), although
these, too, may turn out to be associated with outflows when studied
at high resolution in the mid infrared (e.g., \cite{deBuizer05b}).
The conclusion to be drawn is that a linear distribution of masers
with a velocity gradient does not by itself present convincing
evidence that the masers trace an edge-on disk.  An intriguing variant
is the possibility of methanol masers tracing a face-on disk in
G23.657$-$0.127 (\cite{Bartkiewicz05a}).

There is evidence to support the hypothesis that most methanol masers
are tracing shocked regions, often in the presence of outflows.
Methanol masers appear preferentially near radio sources with a
spectral index indicative of an outflow (\cite{Zapata06}).
Mid-infrared images of some sources indicate that masers are found
along the shocked material on the surface of an outflow cavity
(\cite[De Buizer 2006, 2007]{deBuizer06,deBuizer07}).  In some
sources, methanol masers appear near but offset from UCH~\textsc{ii}
regions, suggesting that they appear in the shocked molecular gas
outside the ionization front, similar to hydroxyl masers
(\cite{Phillips05}).

The 6.7 GHz transition is a popular line for Galactic maser surveys.
Several surveys were reported on during IAU Symposium 206.  Details of
the Arecibo and Parkes multibeam 6.7 GHz surveys can be found in the
section of these proceedings devoted to Galactic maser surveys.
Unsurprisingly, their distribution correlates well with Galactic
structure (\cite{Pestalozzi05}; \cite{Pestalozzi07}).  The 12.2 GHz
line, when it occurs, is almost always weaker than 6.7 GHz emission
(\cite{Blaszkiewicz04}).  Both lines have also been the subject of
monitoring studies (e.g., \cite{Goedhart05a}), which find variability
in a large fraction of sources including periodic variability and a
time delay between features possibly due to light travel time
(\cite{Goedhart03}; \cite{Goedhart05b}).  Based on comparison of
spectra over a period of a decade, the lifetime of an individual
6.7~GHz maser feature is about 150 years (\cite{Ellingsen07}), while
the lifetime of the 6.7~GHz maser phase in a source is a few $\!
\times \, 10^4$ years (\cite{Codella04}; \cite{vanderWalt05}), similar
to the lifetime of the OH maser phase (e.g., \cite{Fish06c}).

Numerous other Class II transitions have been observed.  New maser
sources have been found in rare transitions at 85.5, 86.6, and
107.0~GHz (\cite{Minier02}; \cite{Ellingsen03}) and a
torsionally-excited line at 44.9~GHz (\cite{Voronkov02}).  Several
weak maser lines near 165~GHz have also been detected
(\cite{Salii06}).  Emission in the 19.9~GHz transition is usually weak
and correlates well with 6035~MHz OH masers (\cite{Ellingsen04}).  A
search for 23.1~GHz emission resulted in no new detections beyond the
previously known maser in NGC 6334F (\cite{Cragg04}).  Observations of
these less common methanol maser transitions can help constrain
physical parameters in maser models.  Improved collisional rate data
has also allowed refinement of methanol models, which slightly affects
predicted excitation conditions and brightness temperatures but not
which transitions are expected to produce detectable Class II masers
(\cite{Cragg05}).

\subsection{Hydroxyl (OH)} \label{oh}

Hydroxyl masers are usually studied in sources with associated
UCH~\textsc{ii} regions (e.g., \cite{Fish05a}) but are also
found toward less evolved massive protostellar objects
(\cite{Edris07}).  A different class of OH masers is seen at the ends
of the jet in the W3 TW object (\cite{Argon03}).

Hydroxyl masers around more evolved sources are usually seen in
expansion (sometimes very rapid; see \cite{Stark07}) ahead of the
ionization front of a UCH~\textsc{ii} region (\cite{Fish06c}).
Sometimes the masers appear to trace a molecular disk or torus
(\cite{Slysh02a}; \cite{Hutawarakorn05}; \cite{Edris05};
\cite{Nammahachak06}).  Masers are often seen along arcs or filaments
(\cite{Cohen06}), with extended filamentary emission especially common
at 4.7~GHz (\cite{Palmer03}).

Multitransition overlaps are of special interest because of their
ability to constrain physical conditions in models.  The 4765~MHz line
is observed to be the strongest line of the 6~cm triplet but is
usually only weakly inverted and often spatially extended (e.g.,
\cite{Palmer04}; \cite{HarveySmith05}).  A histogram of 18~cm emission
resembles 4.7~GHz lineshapes in W49A, suggesting that the high-gain
18~cm emission and low-gain 6~cm emission have similar velocity
distributions, even if the 4660~MHz emission is spatially separate
(\cite{Palmer05}).  Much is made of overlaps between 4765 and 1720~MHz
masers (\cite{Palmer03}; \cite[Niezurawska \etal\ 2004,
2005]{Niezurawska04,Niezurawska05}).  It should be noted that 6035~MHz
maser emission correlates more strongly with 4765~MHz than does
1720~MHz, even if the velocities do not always agree (\cite{Dodson02};
\cite{Smits03}).  Masers in the 1720~MHz transition also appear to
correlate with 1665 and 6035~MHz OH masers and 6.7~GHz methanol
masers, at least to arcsecond accuracy (\cite{Caswell04a}).  Masers in
the 6030~MHz transition are almost always accompanied by stronger
emission at 6035~MHz (\cite{Caswell03}), with excellent spatial
coincidence and agreement of magnetic field strengths with each other
and with 1665~MHz masers (\cite{Etoka05}).  Masers in the
highly-excited 13441~MHz transition are rare but are always
accompanied by 6035~MHz masers at the same velocity, although the
intensities in the two transitions do not show a high degree of
correlation (\cite{Baudry02}; \cite{Caswell04c}).  The ground-state,
satellite line transitions at 1612 and 1720~MHz are usually conjugate
with respect to absorption and maser emission (\cite{Szymczak04}).
Sources do exist in which both transitions are inverted, though not in
direct spatial overlap (e.g., \cite{Wright04}).

Magnetic fields as strong as 40~mG are seen in OH masers
(\cite{Slysh06}; \cite{Fish07}).  Magnetic fields are highly ordered
in star forming regions (\cite{Fish06c}) and support pictures in which
the processes of star formation do not tangle field lines
significantly.  While magnetic field strengths are usually stable from
epoch to epoch, monotonic decay of the field in a Zeeman group in
Cep~A continues to be observed (\cite{Bartkiewicz05b}).  High spectral
resolution observations support the conventional assumption that the
Zeeman splitting coefficient appropriate for $\sigma^{\pm 1}$
components should be assumed when measuring magnetic fields at 1612
and 1720~MHz (\cite{Fish06d}).  Linear polarization is of limited
usefulness in determining the full, three-dimensional orientation of
the magnetic field, likely due to a combination of Faraday rotation
and anisotropic magnetohydronamic turbulence (\cite{Watson04};
\cite{Fish06c}).

Extreme variability is occasionally seen in OH.  The 1665~MHz maser in
W75~N briefly flared to nearly 1~kJy to become the brightest OH maser
in the sky (\cite{Alakoz05}).  The 4765~MHz transition is highly
time-variable (\cite{Palmer04}): the maser in Mon R2 flared to nearly
80~Jy before disappearing (\cite{Smits03}) and reappearing
(\cite{Fish06b}).  Short-timescale variability is seen in the
ground-state lines (\cite{Ramachandran06}; also Miller Goss in these
proceedings).

\subsection{Formaldehyde (H$_2$CO)} \label{h2co}

Formaldehyde masers are seen near a handful of several massive YSOs,
with several new detections in recent years (\cite[Araya \etal\ 2005,
2006]{Araya05,Araya06}).  They have both a compact and an extended
component with velocity gradients (\cite{Hoffman03};
\cite{Hoffman07}).  A short-duration flare has been detected toward
one source (\cite{Araya07}).  Further details can be found in the
review by Esteban Araya.

\subsection{Silicon monoxide (SiO)} \label{sio}

While SiO masers are commonly seen in evolved stars, they are rare in
star forming regions.  They are seen in bipolar outflows in W51 IRS 2
and Orion KL source I and appear much closer to the central source
than do water masers (\cite{Eisner02}; \cite{Greenhill04}).  As is the
case in evolved stars, the maser species SiO, water, and OH occur at
progressively larger distances from source I (\cite{Cohen06}).  While
OH masers are ubiquitous throughout Orion, there is a ``zone of
avoidance'' associated with source I in which they do not appear but
SiO and water masers do.  Interestingly, the $v = 1,\, J = 2
\rightarrow 1$ masers are found closer to the protostar in source I
than are the $J = 1 \rightarrow 0$ masers, a finding that is difficult
to understand in the context of SiO maser pumping models
(\cite{Doeleman04}).

\subsection{Other species} \label{other}

Few other new maser species or transitions have been reported in the
literature since the last meeting.  The first $(J,K) = (6,6)$ ammonia
(NH$_3$) maser has been detected centered on a millimeter peak in NGC
6334 I (\cite{Beuther07}).  Weakly inverted acetaldehyde (CH$_3$CHO)
has been detected in the $1_{11} \rightarrow 1_{10}$ transition at
1065.075 MHz toward Sgr B2 (\cite{Chengalur03}).

\section{Multi-species associations} \label{multi}

Methanol, OH, and water masers are frequently in the same source,
although water and methanol masers usually originate in different
regions (\cite{Beuther02}; \cite{Caswell04b}; \cite{Edris05};
\cite{Szymczak05}).  Most 6.7~GHz methanol maser sources have
associated OH masers, almost always at 1665~MHz and frequently at
1667~MHz as well (\cite{Szymczak04}), while the correlation between
6.7~GHz methanol masers and 22~GHz water masers is less strong (e.g.,
\cite{Breen07}).  The distributions of 6.7~GHz methanol masers and
6.0~GHz OH masers in W3(OH) are very similar, although direct overlap
of the two species is rare (\cite{Etoka05}).  Similar phenomena are
also seen between 6.7~GHz methanol and 1.6/4.7~GHz OH masers
(\cite{HarveySmith06}).

All three species correlate more strongly with mid infrared emission
than centimeter or near infrared emission, and all are frequently
found in linear groupings (\cite{deBuizer05}).  However, the
luminosity of water masers correlates less strongly with far infrared
luminosity than is the case for methanol and OH masers, likely because
water masers are not predominantly pumped by infrared photons
(\cite{Szymczak05}), although they may not be pumped entirely by
collisions either (\cite{Liu05}; \cite{Liu07}).  In any case, the
existence of either masers (water and methanol) or outflows towards a
UCH~\textsc{ii} region is an excellent predictor of the other
(\cite{Codella04}), indicating that both masers and outflows are
usually detectable in the UCH~\textsc{ii} phase.

\section{Observational advances} \label{advances}

\subsection{Proper motions and geometric distances} \label{motions}

Evidence in support of the kinematic interpretation of maser motions
continues to pile up, with reports of the persistence of spot shapes
in methanol (\cite{Moscadelli02}) and water masers (\cite{Goddi06b})
and the inferred average overdensity of masers as compared to
non-masing material (\cite{Fish05b}; \cite{Fish06a}).  In addition to
tracing internal source motions, maser proper motions can be used to
obtain geometric parallax distances and measurements of Galactic
rotation.  Recent years have seen this technique used for both
methanol and water masers in W3(OH) to obtain distances accurate to a
few percent (\cite{Xu06}; \cite{Hachisuka06}).  Further details can be
found in proceedings in the Galactic structure session.

\subsection{Spectral resolution} \label{resolution}

Very high spectral line observations at VLBI (very long baseline
interferometry) spatial resolution have been obtained toward masers in
several species in star forming regions.  Detailed line profile
analyses of water masers conclude that the near-Gaussian lineshapes
indicate that they occur in hot ($\sim 1200$~K) gas with small beaming
angles (\cite{Watson02}).  Similar observations of 12.2~GHz methanol
masers (\cite{Moscadelli03}) and the ground-state quartet of OH in
W3(OH) (\cite{Fish06d}) find similarly Gaussian spectral profiles as
well as maser spot positional gradients as a function of velocity
(equivalently, velocity gradients).  The positional gradients show no
clear large-scale spatial organization but have similar magnitudes in
methanol and three of the four OH transitions.  It is possible that
these positional gradients represent turbulent motions on very small
spatial scales.  If so, it is important to understand the
characteristics of the turbulence, since observed maser properties,
including variability, can be highly sensitive to turbulence in the
masing region (\cite{Boger03}; \cite{Sobolev03}; \cite{Silantev06}).
Similar polarization characteristics in Class II methanol features at
different velocities may indicate that velocity gradients induce
velocity redistribution (\cite{Wiesemeyer04}), which may play a
critical role in preventing saturated rebroadening (e.g.,
\cite{Nedoluha88}).

\subsection{Infrared pumping lines} \label{pumping}

Molecular infrared transitions observations, particularly of OH, are
essential for measuring maser pump efficiencies and may help place
observational constraints on the radiative pump cycles of some models
(e.g., \cite{Gray07}).  Archival Infrared Space Observatory data have
been searched for the 34.6 and 53.3~$\mu$m pumping lines of OH with
limited success, due to the low spectral resolution of the instruments
(\cite{He04}; \cite{He05}).  Herschel and SOFIA will have the spectral
resolution and frequency coverage required to observe pumping lines of
OH as well as the critical 560~$\mu$m line of methylidyne (CH).

\section{Further remarks} \label{further}

Two quotes from \cite[De Buizer \etal\ (2005)]{deBuizer05} serve to
summarize the common themes of the observations over the past five
years.  The first is that ``maser emission in general can trace a
variety of phenomena associated with massive stars including shocks,
outflows, infall, and circumstellar disks.  No one maser species is
linked exclusively to one particular process or phenomenon.''  Indeed,
while certain maser species may preferentially turn up in a particular
context (possibly a result of observational biases), the set of all
observations of any one maser species resist being pigeonholed into a
particular phenomenon.  The second quote is that water, OH, and
methanol masers ``do not seem to be associated with different early
evolutionary stages of massive stars.  Instead it appears that they
all trace a variety of stellar phenomena throughout many early stages
of massive stellar evolution.''  As is clear from \S\ref{multi},
different maser species are commonly found together, independent of
the evolutionary stage of the source.  While proposed sequences in
which certain masers turn on before others may be useful for
statistical evaluation of evolutionary phases, important exceptions to
such sequences exist.  Those oddball masers that do not seem to fit
present standard paradigms should be studied in especial detail, since
we cannot predict beforehand what will be thereby learned about their
environment or about maser processes in general.

It was only a few years ago that \cite[Ellingsen (2004)]{Ellingsen04a}
referred to masers as ``the Bart Simpson of star formation research,''
noting that they are ``under-achievers'' in comparison with masers in
other environments due to the lack of sensitive, high resolution
observations at complementary wavelengths.  While this may once have
been true, recent maser observations have made great advancements in
probing a wide range of dynamic structures relevant to star formation.
Maser VLBI allows observations of small- and large-scale morphologies,
magnetic fields, and motions on AU scales and is showing great promise
as a tool to trace Galactic structure.  Maser models for some species
are becoming sufficiently refined to provide good constraints on
physical conditions.  The community is beginning to appreciate the
role of turbulence and the ability to probe its properties using maser
observations.  Synergies with mid infrared instruments have clarified
many of the mysteries of linear structures with velocity gradients.
We have entered the era of greatly improved far infrared
instrumentation, and the ALMA (Atacama Large Millimeter Array) era,
with unprecedented sensitivity and angular resolution at submillimeter
wavelengths, will begin in a few years.  Further advancements in radio
instrumentation, including new space VLBI missions and the SKA (Square
Kilometre Array), will provide even greater insights.  It is perhaps
more correct to state that maser observations are at the vanguard of
star formation research: yesterday's observations can be explained by
complementary data and theory today, and today's observations lay the
groundwork for the breakthroughs that will be achieved in the context
of tomorrow.

\begin{acknowledgments}
The National Radio Astronomy Observatory is a facility of the National
Science Foundation operated under cooperative agreement by Associated
Universities, Inc.
\end{acknowledgments}

\end{document}